\documentclass[prd,twocolumn,aps,amsmath,amssymb,nofootinbib,preprintnumbers]
{revtex4}

\usepackage{graphicx}% Include figure files
\usepackage{dcolumn}% Align table columns on decimal point
\usepackage{bm}% bold math
\usepackage{amsmath}
\usepackage{amsfonts}
\def\ls{\mathrel{\lower4pt\vbox{\lineskip=0pt\baselineskip=0pt
           \hbox{$<$}\hbox{$\sim$}}}}
\def\gs{\mathrel{\lower4pt\vbox{\lineskip=0pt\baselineskip=0pt
           \hbox{$>$}\hbox{$\sim$}}}}
\def\drawbox#1#2{\hrule height#2pt
        \hbox{\vrule width#2pt height#1pt \kern#1pt
              \vrule width#2pt}
              \hrule height#2pt}

\def\Asym#1#2{\vcenter{\vbox{\drawbox{#1}{#2}
              \kern-#2pt       % line up boxes
              \drawbox{#1}{#2}}}}

\newcommand{\beq}{\begin{equation}}
\newcommand{\eeq}{\end{equation}}

\begin{document}

\title{Quasi-thermal Universe and its implications for gravitino produciton, 
baryogenesis and dark matter}

\author{Rouzbeh Allahverdi$^{1}$,}
%\email{averdi@triumf.ca}
\author{Anupam Mazumdar$^{2}$}
% \email{anupamm@nordita.dk}
\affiliation{$^{1}$~Theory Group, TRIUMF, 4004 Wesbrook Mall, Vancouver, BC, 
V6T 2A3, Canada. \\
$^{2}$~NORDITA, Blegdamsvej-17, Copenhagen-2100, Denmark.}

\begin{abstract}
Under general circumstances full thermal equilibrium may not be
established for a long period after perturbative or non-perturbative
decay of the inflaton has completed. One can instead have a
distribution of particles which is in kinetic equilibrium and evolves
adiabatically during this period. Number-violating reactions which are
required to establish chemical equilibrium can become efficient only
at much later times. We highlight some of the striking consequences of
such a quasi-thermal Universe. In particular, thermal gravitino
production yields {\it no bound} on the maximum temperature of the
primordial thermal bath alone. As another consequence, the lower bound
on the mass of the lightest right-handed (s)neutrino from thermal
leptogenesis can be $\gg 10^{9}$~GeV. Depending on the phase, a Wino
or Higgsino considerably lighter than TeV, or a Bino in the bulk
region, can be a viable {\it thermal} dark matter candidate. Finally
the electroweak symmetry may never be restored in the early Universe,
therefore weakening any hopes of realizing a successful electroweak
baryogenesis.
\end{abstract}
\preprint{NORDITA-2005-31, TRI-PP-05-10}
\maketitle

%%%%%%%%%%%%%%%%%%%%%%%%%%%%%%%%%%%%%%%%%%%%%%%%%%%%%%%%%%%%%%%%%%%%%%%%%%%%%%%
\noindent

The detection of the cosmic microwave background anisotropies and the
fluctuations on scales larger than the size of the Hubble radius at
the recombination era~\cite{wmap} is a strong indication of an early
inflationary epoch~\cite{infl}.  An important aspect of inflation is
the reheating, which starts with the inflaton decay and, eventually
results in a thermal bath of elementary particles.  The energy
density, $\rho$, and the number density, $n$, of a distribution of
relativistic particles at a full thermal equilibrium are given by
\beq \label{full}
\rho = g_*\left({\pi^2/ 30}\right) T^4 ~ , 
~ n = g_* \left({\zeta(3)/ \pi^2}\right) T^3\,,
\eeq
where $T$ is the temperature and $g_*$ is the number of degrees of
freedom. The average energy of particles is given by: $\langle E
\rangle = \left(\rho/n \right) \simeq 3T \sim \rho^{1/4}$.

The inflaton decay can be a perturbative~\cite{reheat}, or a
non-perturbative~\cite{preheat} process, while in the models of
running mass inflation the inflaton condensate fragments into unstable
lumps slowly decaying through surface evaporation~\cite{fragment}.  In
all these cases the inflaton decay results in a distribution of
particles which is far from full thermal equilibrium.

Let us denote the inflaton mass (close to the minimum of its
potential) and the decay rate by $m_{\phi}$ and $\Gamma_{\rm d}$
respectively.  At the end of the inflaton decay: $\rho \approx 3
\left(\Gamma_{\rm d} M_{\rm P}\right)^2$, where $M_{\rm P} = 2.4
\times 10^{18}$~GeV is the reduced Planck mass.  When the inflaton
decays perturbatively, $\Gamma_{\rm d} \ll m^2_{\phi}/M_{\rm P}$ and
$\langle E \rangle \approx m_{\phi}/2$, which results in $\langle E
\rangle \gg \rho^{1/4}$. On the other hand for the non-perturbative
inflaton decay as in the case of preheating, $\Gamma_{\rm d} \sim
10^{-2} m_{\phi}$ and $\langle E \rangle \gs 10
m_{\phi}$~\cite{preheat}, implying that $\langle E \rangle \ll
\rho^{1/4}$. In both the cases reaching full equilibrium requires that
$\langle E \rangle$ and, from the conservation of energy, $n$ change
and settle to their respective equilibrium values given by
Eq.~(\ref{full}).

Reactions which conserve the total number of particles re-distribute
the energy and lead to kinetic equilibrium, as well as {\it relative}
chemical equilibrium, among different degrees of freedom. Full thermal
equilibrium also requires number-violating reactions, which are higher
order processes, so that the chemical equilibrium is achieved for all
species. An important point is that in general the inflaton decay,
number-conserving reactions and number-violating reactions occur at
(very) different rates. In consequence, kinetic and chemical
equilibrium can be established at different time scales long after the
inflaton decays.

Indeed this is the case after preheating: the inflaton decay products
(and the fields to which they are coupled) reach kinetic equilibrium
very rapidly, but full equilibrium is established over a much longer
period~\cite{fk}.  In the case of perturbative inflaton decay, the $2
\rightarrow 2$ and $2 \rightarrow 3$ scatterings with gauge boson
exchange have the main role in building kinetic and chemical
equilibrium respectively~\cite{thermalization}. These scatterings are
therefore suppressed if the relevant gauge bosons have large masses in
the early Universe.

This generically happens in supersymmetric theories. Supersymmetry
guarantees flat directions (made up of squark, slepton and Higgs
fields) which acquire a large vacuum expectation value (VEV) during
inflation~\cite{Enqvist}. Flat directions with a large VEV usually
enhance the inflaton decay~\cite{abm}. However, similar to the Higgs 
mechanism, the flat direction
VEV (denoted by $\langle \varphi\rangle$) spontaneously breaks gauge 
symmetries and
induces a mass $m_{\rm g}$ to the gauge bosons, where $m^2_{\rm g} \simeq 
\alpha {\langle\varphi\rangle}^2$ ($\alpha$ is a gauge fine structure 
constant). Note that 
$m_{\rm g}$ has no bearings at low energies/late
times since it vanishes once the flat direction settles to its 
origin~\footnote{Particles which are coupled to the flat direction acquire a 
large mass $\propto \langle \varphi \rangle$, and hence promptly decay to 
massless particles (i.e. the ones which are not coupled to the flat 
direction). Such a decay can increase the number of particles by a factor of 
$2$ or $3$, depending on whether the dominant channel is $1 \rightarrow 2$ 
or $1 \rightarrow 3$ decay. However this alone is too small to lead to full 
chemical equilibrium, as it needs the number of particles to increase by a 
huge factor (typically $> 10^3$) after perturbative inflaton decay.}.

Kinetic equilibrium is usually built at a rate $\Gamma_{\rm kin} \sim
\alpha^2 n/m^2_{\rm g}$, while full equilibrium is established at a
rate $\Gamma_{\rm thr} \sim \alpha^3 n/m^2_{\rm g}$ ($n$ is the number
density of inflaton decay products)~\cite{thermalization}. For
sufficiently large $\langle\varphi\rangle\sim {\cal
O}(10^{13}-10^{18})$~GeV and $\alpha \sim 10^{-2}$, it turns out that
$\Gamma_{\rm thr} \ll \Gamma_{\rm d}$. This substantially delays
thermalization and results in a (very) low reheat temperatures (for
example, $T_{\rm R} \sim {\cal O}({\rm TeV})$ if $\langle \varphi
\rangle \sim 10^{18}$ GeV)~\cite{amm}.

Therefore under general circumstances the Universe enters a
quasi-thermal phase.  During this phase the comoving number density and 
comoving average energy of particles remain constant. It is then paramount 
that the Universe reaches
full equilibrium before nucleosynthesis~\cite{BBN}, i.e., $T_{\rm R}
\gs {\cal O}({\rm MeV})$, which is known to tolerate hardly any
departure from thermal equilibrium. The aim of this paper is to show
striking consequences of such a quasi-thermal epoch, such as particle
production and phase transitions.

During the quasi-thermal phase we can parameterize the energy density
and the number density of the $i-$th degree of freedom as:
\beq \label{kinetic}
\rho_i = C_i ~A_i \left({\pi^2 / 30}\right) T^4 ~ , ~ 
n_i = A_i \left({\zeta(3) /\pi^2}\right) T^3\,.
\eeq
where $C_i,A_i \neq 1$ indicate deviation from full equilibrium (i.e.,
$C_i = A_i = 1$ for bosons and $A_i = 3/4,C_i = 7/6$ for fermions).
Note that relative chemical equilibrium among different species may be
established much later than kinetic equilibrium (as it happens after
the perturbative inflaton decay~\cite{amm}), and hence $A_i$ have
different values in general.  If $A_i < 1$, the number of particles is
smaller than that in full thermal equilibrium. We can then assign a
negative chemical potential $\mu_i$ to the $i-$th particle where $A_i
= {\rm exp}(\mu_i/T)$, and $C_i \simeq \langle E\rangle/3T \approx
1$. This is a typical situation for the perturbative inflaton 
decay~\footnote{Since in this case 
particles are produced in one-particle decay of the inflaton, the distribution 
is peaked around the average energy. The $2 \rightarrow 2$ scatterings (which 
become efficient shortly before complete thermalization) smooth out the 
distribution. This implies that the reheat plasma is close to kinetic 
equilibrium but grossly deviated from chemical equilibrium, hence 
is in a quasi-thermal state.}.

The situation will be more complicated for $A_i > 1$, because the
number of particles is (much) larger than the full equilibrium value,
thus implying a positive chemical potential. For bosons with a typical
mass $m_i < T$, we have $\mu_i \leq m_i$. This implies population of
the ground state and renders large occupation numbers. Once $n_i \gg
T^4/m_i$, the ground state dominates both the number and energy
densities. We can then use Eq.~(\ref{kinetic}) where $A_i \gg 1$ and
$C_i \sim m_i/T \ll 1$. Similarly, for a degenerate Fermi gas with
$\mu_i \gg T$, Eq.~(\ref{kinetic}) is applicable with $A_i \sim
\left(\mu_i/T \right)^3 \gg 1$ and $C_i \sim \mu_i/T \gg 1$. At the
end of preheating the average kinetic energy and the (effective) mass
of particles in the plasma are approximately
equal~\cite{preheat}. This implies that $m_i \sim T$ and $C_i \sim
1$. Therefore, in cases of physical interest, we can take $C_i \simeq
1$ in Eq.~(\ref{kinetic}).

The quasi-thermal phase then lasts until $H \simeq \Gamma_{\rm thr}$,
where $\Gamma_{\rm thr}$ is the rate for number-violating reactions.
During this period $A_i$ remain practically constant and $T \propto R^{-1},~n_i
\propto R^{-3}, ~\rho_i \propto R^{-4}$ (with $R$ being the scale
factor of the Universe). This implies an adiabatic evolution of the
plasma during which
\beq
\label{req2}
H \simeq A^{1/2} \left(T^2/3M_{\rm P}\right) ~ ~ , ~ ~ 
A \equiv \sum_{i}{A_i}\,.
\eeq
At full equilibrium $A = g_*$, where $g_* \simeq 200$ in the minimal
supersymmetric standard model (MSSM). The plasma has its maximum
temperature $T_{\rm max}$ at the end of the inflaton decay, i.e., when
$H \simeq \Gamma_{\rm d}$. From Eq.~(\ref{req2}) we obtain,
\beq \label{a}
A \sim 10 \left(\Gamma^2_{\rm d} M^2_{\rm P}/T^4_{\rm max}\right)\,. 
\eeq
Note that $n_i$ and $A_i$ are determined by the branching ratio for
the inflaton decay to the $i-$th degree of freedom. Perturbative
inflaton decay generically yields $A_i \ll 1$ (thus $A \ll g_*$).  For
example, consider a gravitationally decaying inflaton $\Gamma_{\rm d}
\sim m^3_{\phi}/M^2_{\rm P}$ with $m_{\phi} = 10^{13}$ GeV. This leads
to $T_{\rm max} \sim 10^{12}$ GeV (note that $\langle E \rangle \simeq
3 T_{\rm max} \approx m_{\phi}/2$) and $A \sim 10^{-8}$. On the other
hand, preheating results in the opposite situation where $A_i \gg 1$
(and $A \gg g_*$).  As mentioned earlier $T_{\rm max} \sim 10
m_{\phi}$ and $\Gamma_{\rm d} \sim 10^{-2} m_{\phi}$ in this
case~\cite{preheat}. Then, for $m_{\phi} = 10^{13}$ GeV, this yields
$A \sim 10^4$.

Once $H \simeq \Gamma_{\rm thr}$, at which time the temperature has
dropped to $T_{\rm F}$, the number-violating reactions become
important and lead to full equilibrium (i.e., $A_i \simeq 1$) very
quickly.

We define the reheat temperature $T_{\rm R}$ as the temperature of the
distribution {\it after} the establishment of full thermal
equilibrium, therefore $T_{\rm R} \simeq \left(\Gamma_{\rm thr} M_{\rm
P}\right)^{1/2}$. In general we can have $\Gamma_{\rm thr} \ll
\Gamma_{\rm d}$, and hence $T_{\rm R} \ll \left(\Gamma_{\rm d} M_{\rm
P}\right)^{1/2}$, which is often used in the
literature~\cite{infl}. The conservation of energy implies that
$T_{\rm R} = \left(A/g_* \right)^{1/4} T_{\rm F}$. The entropy
density, which should be used to normalize a given number density, is
given by
\beq
\label{req1}
s = A^{3/4} g^{1/4}_* \left({2 \pi^2 / 45}\right) ~ T^3_{\rm F}\,.
\eeq
Let us now consider the production of a (stable or long-lived)
particle $\chi$ with mass $m_{\chi}$ in the quasi-thermal phase. If
the couplings of $\chi$ to the fields in the thermal bath are
sufficiently small, it will never reach equilibrium. Then its
abundance will be subject to only production and redshift, and we will
have
\beq \label{chipro} 
\dot{n}_{\chi} + 3 H n_{\chi} = \sum_{ij}{\left[A_i A_j \left({\zeta(3)/
\pi^2}\right)^2 \langle \sigma^{ij}_{\chi} v_{\rm rel} \rangle ~ T^6\right]}\,.
\eeq
Here $\sigma^{ij}_{\chi}$ denotes the cross-section for production of
$\chi$ from scatterings of the $i-$th and $j-$th particles and the sum
is over all particles which participate in $\chi$ production. Also
$\langle \rangle$ and the dot denote thermal averaging and proper time
derivative, respectively, and we have used Eq.~(\ref{kinetic}). Note
that $\chi$ production is kinematically possible in the interval
$m_{\chi} \ls T \leq T_{\rm max}$.

One comment is in order before proceeding further. In supersymmetric theories 
the last stage of inflaton 
decay which reheats the Universe is generically perturbative and results in 
$A_i < 1$~\cite{amm}. In this (physically relevant) case Eq.~(\ref{chipro}) 
can be relibaly used to calculate $\chi$ production in the 
quasi-thermal phase. 
However the situation will be more subtle when $A_i \gg 1$. In this 
case we have a dense plasma where the occupation number of particles is large 
(i.e. $\gg 1$ for bosons and $\simeq 1$ for fermions). 
Such large occupation numbers, besides final state effects which enhance or 
suppress the 
interaction rates, result in large induced masses and significant off-shell 
effects. Therefore the situation is considerably more complicated and a proper 
(non-equilibrium) quantum field theory treatment~\cite{bs} will be required in 
this case. Thus the following expressions of particle abundances must be taken 
with care and considered only as order of magnitude estimates when $A_i \gg 
1$. 
     
%%%%%%%%%%%%%%%%%%%%%%%%%%%%%%%%%%%%%%%%%%%%%%%%%%%%%%%%%%%

\vskip5pt {\it Gravitino-} If $\chi$ is the gravitino,
$\sigma^{ij}_{\chi} \sim \alpha/M^2_{\rm P}$. In general, an exact
calculation of the gravitino abundance is complicated since various
scattering processes have different cross-sections (similar argument
will be followed for the rest of the paper). The result now depends on
the values of $A_i$, and hence is model-dependent. The maximum
abundance is obtained when the plasma mainly consists of gluons as
$gluon + gluon \rightarrow gluino + gravitino$ scattering has the
largest cross-section.  Following the standard
calculations~\cite{thermal}, and with the help of
Eqs.~(\ref{req2},\ref{req1}), we then find an upper bound (for a more detailed 
calculation see~\cite{amm}) 
\beq \label{gravitino1}
\left({n_{3/2}/s}\right) \sim 
A^{3/4} \left({T_{\rm max}/ 10^{10}~{\rm GeV}}\right) ~ \times 10^{-12}. 
\eeq
in the case of MSSM.  For a TeV scale gravitino the success of BBN
requires that~\cite{bbn}
\beq
\label{const1}
T_{\rm max} \leq A^{-3/4} \times 10^{10}~{\rm GeV}\,. 
\eeq
This bound depends crucially on the parameter $A$. To illustrate, let
us again consider the example of a gravitationally decaying inflaton
for which $A \sim 10^{-8}$. It is interesting that despite having
$T_{\rm max} \sim 10^{12}$ GeV, the constraint in Eq.~(\ref{const1})
is comfortably satisfied.  Note that now the bounds from thermal
gravitino production on the inflaton sector can be substantially
relaxed. Therefore within the regime of perturbative decay there is
{\it no strict bound} on maximum temperature of the Universe.

On the other hand in the case of preheating (where $A \gg 1$), the
constraint from the overproduction of gravitinos become: $m_{\phi} <
T_{\rm max} \ll 10^{10}$~GeV. For such a small inflaton mass chaotic
type inflation will be a disaster from the point of view of matching
the observed temperature anisotropy, one would have to resort for
alternative scenarios as discussed in Ref.~\cite{Kari}. Otherwise one
would require a larger gravitino mass which would decay early enough
as discussed in Ref.~\cite{Rouzbeh}.  In passing we note that
gravitinos are also created {\it during} preheating from vacuum
fluctuations~\cite{Mazumdar}. This non-perturbative production has
nothing to do with thermal scatterings, and hence is not modified in
the presence of a quasi-thermal phase.

%%%%%%%%%%%%%%%%%%%%%%%%%%%%%%%%%%%%%%%%%%%%%%%%%%%%%

\vskip5pt {\it Right-handed (s)neutrino-} If $\chi$ is the lightest
right-handed (RH) (s)neutrino, its decay can generate the observed
baryon asymmetry of the Universe. In the thermal leptogenesis scenario
these (s)neutrinos are mainly produced from scatterings of the
left-handed (LH) (s)leptons and the top (s)quarks~\cite{lepto}.  The
exact result is again model-dependent. However, the generated baryon
asymmetry in the quasi-thermal phase is maximum when the plasma mainly
consists of the LH (s)leptons and the top (s)quarks. Applying
Eqs.~(\ref{req2},\ref{req1}) to the standard
calculations~\cite{lepto}, then results in an upper bound (for a more detailed 
calculation see~\cite{amm}):
\beq \label{lepto}
\eta_{\rm B} \equiv {n_{\rm B}/ s} \sim A^{3/4} \kappa 
\left({M_1 / 10^{9} ~ {\rm GeV}}\right) ~ \times 3 \cdot 10^{-10}\,,
\eeq
where $M_1$ is the mass of the lightest RH (s)neutrino, and $\kappa
\leq 10^{-1}$ is the efficiency factor accounting for the processes
involving the RH states. Generating sufficient asymmetry $\eta_{\rm B}
\simeq 0.9 \times 10^{-10}$ then sets a lower bound on $A^{3/4} M_1$
instead of $M_1$ alone:
\beq \label{const2}
M_1 \geq A^{-3/4} \times 10^9~{\rm GeV}\,.
\eeq
For a perturbative inflaton decay ($A \ll 1$), this results in $M_1
\gg 10^9$~GeV. However, $M_1 \ls 3 T_{\rm max} \approx m_{\phi}/2$ is
also required in order to kinematically produce the RH (s)neutrinos,
where $m_{\phi}\leq 10^{13}$~GeV. This, together with
Eq.~(\ref{const2}), leads to a lower bound $A > 10^{-5}$. This implies
that for a gravitationally decaying inflaton $A \sim 10^{-8}$ is too
small to allow successful leptogenesis.

In the case of preheating, where $A \gg 1$, Eq.~(\ref{const2}) then
implies that successful leptogenesis is possible for much lighter RH
(s)neutrinos $M_1 \ll 10^9$~GeV. One point to note is that the
gravitino abundance has the same dependence on $A$ as the baryon
asymmetry, see Eq.~(\ref{gravitino1},\ref{lepto}). Therefore thermal
leptogenesis remains marginally compatible with thermal gravitino
production in a quasi-thermal Universe as well.

%%%%%%%%%%%%%%%%%%%%%%%%%%%%%%%%%%%%%%%%%%%%%%%%%%%

\vskip5pt {\it Thermal dark matter-} If $\chi$ is one of the particles
in the thermal bath, its distribution will follow
Eq.~(\ref{kinetic}). This happens, for example, when $\chi$ is the
lightest supersymmetric particle (LSP). If $T_{\rm F} < m_{\chi}$, the
$\chi$ quanta become non-relativistic before full thermal equilibrium
is established. Again assuming that the plasma mainly consists of
$\chi$, its equilibrium number density and energy density are in this
case given by:
\beq \label{chi}
n_{\chi} \sim A \left({m_{\chi} T/ 2 \pi}\right)^{3/2} 
{\rm exp}\left(-{m_{\chi}/ T}\right)~,~\rho_{\chi} = m_{\chi} n_{\chi}\,. 
\eeq
The annihilation of $\chi$ quanta, happening at a rate $\Gamma_{\rm
ann} = \left(c_{\chi}/m^2_{\chi}\right) n_{\chi}$, keeps them in
equilibrium until the time when $\Gamma_{\rm ann} < H$. This occurs at
$T = T_{\rm f}$, where the {\it freeze-out} temperature, $T_{\rm f}$,
follows:
\beq \label{freeze1}
{m_{\chi}/ T_{\rm f}} = x_{\rm f}~,~x_{\rm f} \simeq 
0.5 {\rm ln} A + {\rm ln} \left[0.2 ~ c_{\chi} 
\left({M_{\rm P}/m_{\chi}}\right) \right]\,.  
\eeq
The entropy release upon full thermalization, given by
Eq.~(\ref{req1}), results in the final abundance
\beq \label{const3}      
{n_{\chi}/ s} \sim A^{-1/4} x_{\rm f} \left({0.2/c_{\chi}}\right) 
\left({m_{\chi}/ M_{\rm P}}\right)\,. 
\eeq
We remind that in the case of full equilibrium, the unitarity bound
implies that $c_{\chi} \leq 4 \pi$, and hence the thermal abundance
exceeds the dark matter limit $\left(n_{\chi}/s \right) \leq 5 \times
10^{-10} \left(1~{\rm GeV}/m_{\chi}\right)$ if $m_{\chi} >
100$~TeV~\cite{gk}. Freeze-out during the quasi-thermal phase, see
Eq.~(\ref{const3}), constrains $A^{-1/4} m^2_{\chi}$ instead. Thermal
dark matter with $m_{\chi} > 100$~TeV will be now viable if $A \gg
1$. On the other hand, the upper bound on $m_{\chi}$ will be
strengthened if $A \ll 1$.

There are particularly interesting consequences for the LSP dark
matter~\cite{dark}. In full equilibrium, viable Wino-or Higgsino-like
dark matter requires that $m_{\chi} > 1$~TeV. This is considered too
heavy if weak scale supersymmetry is the solution to the hierarchy
problem. Freeze-out in the quasi-thermal phase with $A \ll 1$ allows
$m_{\chi}$ to be (much) closer to the ALEPH bound on the lightest
neutralino mass ($54$~GeV at $95\%$ c.l.~\cite{ALEPH}). If full
equilibrium is assumed, a Bino-like LSP overcloses the Universe in
most parts of the parameter space due to its much smaller annihilation
cross-section. However, see Eq.~(\ref{const3}), $A \gg 1$ can
significantly extend the acceptable part of the parameter space to the
bulk region as it lowers the LSP abundance.

%%%%%%%%%%%%%%%%%%%%%%%%%%%%%%%%%%%%%%%%%%%%%%%%%%%%%%%%%%%%%%%%%%%%
\vskip5pt

{\it Electroweak phase transition-} Finite temperature effects make a
typical contribution $n/T$ to the $({\rm mass})^2$ of every scalar
field which is coupled to the thermal bath. In full equilibrium,
Eq.~(\ref{full}), thermal corrections are $\sim + T^2$. Therefore, at
sufficiently high temperatures, they can lead to restoration of a
broken symmetry~\cite{symm}. Thermal masses can also affect the
dynamics of cosmological Bose condensates by triggering their early
oscillations~\cite{early}.

For a quasi-thermal distribution, Eq.~(\ref{kinetic}), thermal
corrections are $\sim A T^2$ and depending on $A$ there will be
different situations. As a specific example, consider the electroweak
symmetry breaking happening at a scale $m_{\rm EW} \sim 100$~GeV. This
implies that the electroweak phase transition occurs at a temperature
$T_{\rm cr} \sim A^{-1/2} m_{\rm EW}$.

If $A < \left(m_{\rm EW}/T_{\rm max}\right)^2$, then $T_{\rm cr} >
T_{\rm max}$ and the electroweak symmetry will be broken after
inflation. If full thermalization occurs very late, such that $T_{\rm
R} < 100$~GeV, it will never be restored and the electroweak
baryogenesis will not be possible.

On the other hand, $T_{\rm cr} \ll 100$~GeV if $A \gg 1$. Such a late
phase transition will also be problematic for the electroweak
baryogenesis, since sphaleron transitions are not active at that time.
Note that $A \gg 1$ can also result in a large mass $m_W \gg T$ for
the electroweak gauge bosons at $T \gg T_{\rm cr}$. The sphaleron
interaction rate, which is $\propto {\rm exp}\left(-m_{W}/T \right)$,
will then be exponentially suppressed even {\it before} the
electroweak phase transition.

%%%%%%%%%%%%%%%%%%%%%%%%%%%%%%%%%%%%%%%%%%%%%%%%%%%
\vskip5pt 

{\it Discussion-} The most important message of this paper is that
thermal particle production does not solely depend on the temperature
of the Universe. The reason is that under general circumstances the
early Universe can enter a quasi-thermal phase after the inflaton
decay.  This epoch, during which the Universe is far from full thermal
equilibrium, can last very long and include a wide range of
temperatures, particularly those which are relevant to the production
of gravitinos, baryon asymmetry or dark matter.

In particular, we showed that thermal gravitino production yields {\it
no bound} on the maximum temperature of the Universe alone.  The lower
bound on the lightest RH (s)neutrino from thermal leptogenesis can
also be very different from $10^9$ GeV.  Depending on the thermal
history, a Wino- or Higgsino-like LSP much lighter than TeV, or a
Bino-like LSP in the bulk region, can be a viable {\it thermal} dark
matter candidate. A detailed account of thermalization in
supersymmetric models and its cosmological consequences will be given
in a separate publication~\cite{amm}.

%%%%%%%%%%%%%%%%%%%%%%%%%%%%%%%%%%%%%%%%%%%%%%%%%%%%%%%%%%%%%%%%%%%%%%%%%%%%%%%
{\it Acknowledgments-} The authors are thankful to C. Burgess,
B. Dutta, P. Gondolo, Y. Grossman, A. Jokinen, L. Kofman, M. Peloso,
A. Masiero, T. Moroi, T. Mult\"amaki, A. Rajaraman and M. Schelke for
helpful discussions. The work of R.A. is supported by the National
Sciences and Engineering Research Council of Canada.

%%%%%%%%%%%%%%%%%%%%%%%%%%%%%%%%%%%%%%%%%%%%%%%%%%%%%%%%%%%%%%%%%%%%%%%%%%%%%%%

\end{document}